\documentclass[a4paper,oneside,12pt]{article}%
\usepackage[a4paper]{geometry}
\usepackage{graphicx}
\usepackage{amsmath}
\usepackage{amsfonts}
\usepackage{amssymb}%
\providecommand{\U}[1]{\protect\rule{.1in}{.1in}}

\begin{document}

\title{Waves, analytical signals, and some postulates of quantum theory}
\author{Miguel A. Muriel
\and {\small Dept. de Tecnolog\'{\i}a Fot\'{o}nica, Universidad Polit\'{e}cnica de
Madrid,}
\and {\small ETS Ingenieros de Telecomunicaci\'{o}n, Ciudad Universitaria, E-28040
Madrid (Spain).\medskip}
\and F. Javier Fraile-Pel\'{a}ez
\and {\small Dept. de Teor\'{\i}a de la Se\~{n}al y Comunicaciones, Universidad de
Vigo,}
\and {\small ETS Ingenieros de Telecomunicaci\'{o}n, Campus Universitario E-36310
Vigo (Spain).}}
\date{}
\maketitle

\begin{abstract}
In this paper we apply the formalism of the analytical signal theory to the
Schr\"{o}dinger wavefunction. Making use exclusively of the wave-particle
duality and the principle of relativistic covariance, we actually derive the
form of the quantum energy and momentum operators for a single nonrelativistic
particle. Without using any more quantum postulates, and employing the
formalism of the characteristic function, we also derive the
quantum-mechanical prescription for the measurement probability in such
cases.\bigskip

03.65.Ta, 03.65.Pm, 02.30.Nw\newpage

\end{abstract}

\section{Introduction\label{s1}}

This paper presents a derivation of two measurement postulates of Quantum
mechanics using exclusively de Bloglie's original relationship and Planck's
formula as the only previous quantum postulates assumed. Namely, the
associations $E\rightarrow-i\hbar\partial/\partial t$ and $p-i\hbar
\partial/\partial x$ for (nonrelativistic) particles are proved for
one-particle mechanical systems, as well as the postulated formula of the
measurement probability for the two observables, both with discrete and
continuous spectra.

Starting from general wave equations, a theory is constructed based on the
envelope of the space-time analytical signal of a certain\ real-valued
\textquotedblleft auxiliary\textquotedblright\ wavefunction, modulated in a
latent way by its relativistic rest energy. Apart from the relativistic
covariance, in this paper we will only use the principle of wave-particle
duality as the starting postulate, which is summarized by the elementary relations%

\begin{align}
E  &  =\hbar\omega\label{Eh}\\
\mathbf{p}  &  =\hbar\mathbf{k.} \label{ph}%
\end{align}
Actually, the Planck-Einstein relationship (\ref{Eh}) can even be derived on
purely relativistic grounds, with the only additional assumption of the
existence of an energy quantum, but without the need of postulating the
specific value $\hbar\omega.$ \cite{Field}

It is usual in Optics to employ mathematical tools from the Signal Theory
field, where the Fourier transforms and the representations of dual
transformed spaces (space - time - spatial frequency - temporal frequency) are
customarily used. The analytical signals and their paraxial equations allow to
undertake the study of signals of extremely high frequencies like the optical
fields. However, the mathematical formalism of the analytical signal tends to
be handled in a hardly rigorous way in the literature, sometimes even leading
to major errors \cite{Javier}. We anticipate that a careful, accurate use of
the analytical signal and related concepts will be \emph{essential} in this
paper. The notational rigor should not be underestimated in what follows.

Section \ref{s2} summarizes the necessary formalism of the Fourier transforms
and analytical signals. In Section \ref{s3and4}, the Klein-Gordon and
Schr\"{o}dinger equations for a particle in a potential are reviewed with the
new notation. Sections \ref{s5} and \ref{s6} present the derivation of the
quantum-mechanical formula for the average energy and momentum of a
non-relativistic particle. Special attention is paid to the energy stationary
states in Section \ref{sest}. With these results, the postulate of the
measurement probability is derived in Section \ref{s7}. The conclusions of the
work are summarized in Section \ref{s8}.

\section{Analytical signals and Fourier transforms in time and space
domains\label{s2}}

We need to review briefly a few concepts of the analytical signal theory, as
well as introduce the notation we will use. Consider a space-time scalar
\textquotedblleft wavefunction\textquotedblright\ $\psi(\mathbf{r},t)$. For
reasons that will become clear later, $\psi(\mathbf{r},t)$ is considered to be
a \emph{real-valued} wavefunction; consequently, $\psi(\mathbf{r},t)$ does
\emph{not} coincide with the customary \textquotedblleft
wavefunction\textquotedblright\ used in quantum mechanics.

In the base of complex plane waves $\exp[i(\mathbf{k}\cdot\mathbf{r}-\omega
t)],$ $\psi(\mathbf{r},t)$ admits four representations related through
standard Fourier transforms (FT) and their inverses:%

\begin{equation}%
\begin{array}
[c]{ccc}%
\psi(\mathbf{r},t) & \longleftrightarrow & \Psi(\mathbf{r},\omega)\\
\updownarrow &  & \updownarrow\\
\tilde{\psi}(\mathbf{k},t) & \longleftrightarrow & \tilde{\Psi}(\mathbf{k}%
,\omega).
\end{array}
\end{equation}

We will denote its $\omega$-Fourier (time) transform by an uppercase symbol:
$\Psi(\mathbf{r},\omega);$ its $k$-Fourier (space) transform by a tilde:
$\tilde{\psi}(\mathbf{k},t),$ and its $(\mathbf{k},\omega)$-Fourier
(space-time) transform by $\tilde{\Psi}(\mathbf{k},\omega).$ We have%

\begin{equation}
\tilde{\Psi}(\mathbf{k},\omega)=\int\int\int\int_{-\infty}^{\infty}%
\psi(\mathbf{r},t)\,e^{-i(\mathbf{k}\cdot\mathbf{r}-\omega t)}d^{3}r\,dt.
\end{equation}

We will make use throughout of some basic properties of the FT such as%

\begin{equation}%
\begin{array}
[c]{lcl}%
\partial\psi(\mathbf{r},t)/\partial t & \longleftrightarrow & -i\omega
\Psi(\mathbf{r},\omega)\\
\nabla\psi(\mathbf{r},t) & \longleftrightarrow & i\mathbf{k}\,\tilde{\psi
}(\mathbf{k},t).
\end{array}
\end{equation}
\qquad\qquad

In general, if we consider $\psi(\mathbf{r},t)$ as an arbitrary time function,
its \emph{time analytical signal} or \emph{time complex pre-envelope}, denoted
by $\psi_{+}(\mathbf{r},t)$, is defined as the function whose $\omega$-FT is
as follows:%

\begin{equation}
\Psi_{+}(\mathbf{r},\omega)=\left\{
\begin{array}
[c]{ll}%
2\Psi(\mathbf{r},\omega), & \omega>0\\
\Psi(\mathbf{r},0), & \omega=0\\
0, & \omega<0.
\end{array}
\right.
\end{equation}
So, we consistently write%

\begin{equation}
\psi_{+}(\mathbf{r},t){\small \longleftrightarrow}\Psi_{+}(\mathbf{r},\omega).
\end{equation}

Thus, $\psi_{+}(\mathbf{r},t)$ essentially contains only the positive part of
the $\omega$-spectrum of $\psi(\mathbf{r},t).$ As it is well known, $\psi
_{+}(\mathbf{r},t)=\psi(\mathbf{r},t)+i\check{\psi}(\mathbf{r},t),$ where
$\check{\psi}(\mathbf{r},t)$ is the Hilbert transform of $\psi(\mathbf{r},t).$
[The simplest example is given by the complex exponential: $\exp(i\omega
t)=\cos(\omega t)+i\sin(\omega t).$]

Let us note that if $\psi(\mathbf{r},t)$ is real, $\check{\psi}(\mathbf{r},t)
$ is real too, so we can write%

\begin{equation}
\psi(\mathbf{r},t)=\operatorname{Re}[\psi_{+}(\mathbf{r},t)]\equiv
\operatorname{Re}[\overset{_{\frown}}{\psi}(\mathbf{r},t)\exp(-i\omega_{c}t)].
\label{r5}%
\end{equation}
In analytical signal theory, the function $\overset{_{\frown}}{\psi
}(\mathbf{r},t)$ is the so-called \emph{time complex envelope}. The angular
frequency $\omega_{c}$ is in principle an arbitrary parameter but, in optics
or signal theory in general, the decomposition in (\ref{r5}) usually turns out
to be useful when $\psi(\mathbf{r},t)$ is a bandpass signal and $\omega_{c}$
is precisely chosen to be its carrier frequency. The physical meaning of
$\omega_{c}$ in our present case will be explained in Section \ref{s3and4}.

For the sake of notational simplicity and clarity of the exposition, we will
assume hereafter one-dimensional wave propagation in space (say in the $x$
direction), the generalization to three dimensions being obvious and omitted.

Now a separation between positive and negative spectral components can also be
done in the spatial domain. Rather than $+$ and $-,$ we will use the notation
$>$ and $<$ for the $k$-spectrum. For instance, the \textquotedblleft spatial
analytical signal\textquotedblright\ is the function containing only the
spatial frequencies $k>0$. We denote it by $\tilde{\psi}_{%
{\scriptscriptstyle{\rm>}}%
}$, so that%

\begin{equation}
\tilde{\psi}_{%
{\scriptscriptstyle{\rm>}}%
}(k,t)=\left\{
\begin{array}
[c]{ll}%
2\tilde{\psi}(k,t), & k>0\\
\tilde{\psi}(0,t), & k=0\\
0, & k<0,
\end{array}
\right.
\end{equation}
and we can write%

\begin{equation}
\psi_{%
{\scriptscriptstyle{\rm>}}%
}(x,t)\longleftrightarrow\tilde{\psi}_{%
{\scriptscriptstyle{\rm>}}%
}(k,t).
\end{equation}
If $\psi(x,t)$ is real, we also obtain, analogous to the first equality of Eq.
(\ref{r5}),%

\begin{equation}
\psi(x,t)=\operatorname{Re}[\psi_{%
{\scriptscriptstyle{\rm>}}%
}(x,t)].
\end{equation}
In quantum electrodynamics, the $k$-spectrum decomposition is frequently used
---generally as a discrete sum of traveling or stationary modes--- to quantify
the electromagnetic field in the Heisenberg picture (see for example
\cite{CohenT}, \cite{Loudon}); this has nothing to do with its purpose in the
present work.

To keep the space-time analogy complete, we might carry out the spatial
Hilbert carrier and complex envelope decomposition analogous to (\ref{r5}%
).\ However, as we will see later on, these concepts are unnecessary in the
$k$\ domain, so we do not need to elaborate on them.

Finally, we can write%

\begin{equation}
\tilde{\Psi}_{%
{\scriptscriptstyle{\rm>+}}%
}(k,\omega)=\left\{
\begin{array}
[c]{ll}%
2\tilde{\Psi}(k,\omega), & k>0,\,\omega>0\\
\tilde{\Psi}(0,0), & k=0,\,\omega=0\\
0, & k<0,\,\omega<0,
\end{array}
\right.
\end{equation}
so that $\psi_{%
{\scriptscriptstyle{\rm>+}}%
}(x,t)%
\begin{array}
[c]{c}%
_{\longleftrightarrow}\\
^{\longleftrightarrow}%
\end{array}
\tilde{\Psi}_{%
{\scriptscriptstyle{\rm>+}}%
}(k,\omega),$ where the two arrows denote double Fourier transformation (space
and time).

\section{Schr\"{o}dinger equation for $\overset{_{\frown}}{\psi}_{%
{\scriptscriptstyle{\rm>}}%
}$\label{s3and4}}

The space-time planes of the aforementioned waves move with a constant phase
velocity, $v_{p},$ determined by $v_{p}=\omega/k.$ Thus,%
\begin{equation}
v_{p}=\frac{\omega}{k}=\frac{E}{p}, \label{3}%
\end{equation}
where use has been made of the quantum relationships (\ref{Eh}) and (\ref{ph})
in the second equality. Hence,%

\begin{equation}
k^{2}-\omega^{2}/v_{p}^{2}=0. \label{01}%
\end{equation}
This identity remains valid if multiplied for any well-behaved (complex, in
general) function of $k$ and $\omega.$ We multiply Eq. (\ref{01}) precisely by
the function $\tilde{\Psi}_{%
{\scriptscriptstyle{\rm>+}}%
}(k,\omega),$ for reasons that will become clear later. We thus obtain the
equation $(k^{2}-\omega^{2}/v_{p}^{2})\tilde{\Psi}_{%
{\scriptscriptstyle{\rm>+}}%
}(k,\omega)=0.$ Taking its inverse $(k,\omega)$-FT, a generic homogeneous wave
equation is obtained for the doubly-analytical function $\psi_{%
{\scriptscriptstyle{\rm>+}}%
}(x,t)$:%

\begin{equation}
\frac{\partial^{2}\psi_{%
{\scriptscriptstyle{\rm>+}}%
}(x,t)}{\partial x^{2}}-\frac{1}{v_{p}^{2}}\frac{\partial\psi_{%
{\scriptscriptstyle{\rm>+}}%
}(x,t)}{\partial t^{2}}=0. \label{1}%
\end{equation}
Indeed, a necessary (but not sufficient) condition for any such equation to
meet relativistic covariance (with scalar coefficients) is to be second order
both in space and time.

We must now cast Eq. (\ref{1}) into its specific form. Recall that, for a free
particle of rest mass (or simply \textquotedblleft mass\textquotedblright%
\ \cite{Adler}) $m$ moving at a speed $v,$ the relativistic energy and the
spatial part of the four-momentum, as measured in an inertial frame, are given by%

\begin{equation}
E=mc^{2}(1-v^{2}/c^{2})^{-1/2};\qquad p=mv(1-v^{2}/c^{2})^{-1/2}=c^{-1}%
(E^{2}-m^{2}c^{4})^{1/2}. \label{4}%
\end{equation}

Using Eqs. (\ref{4}) and (\ref{Eh}), the equation $k^{2}=\omega^{2}%
/c^{2}-m^{2}c^{2}/\hbar^{2}$ follows from Eq. (\ref{01}). Again, multiplying
this equality through by $\tilde{\Psi}_{%
{\scriptscriptstyle{\rm>+}}%
}(k,\omega)$ and taking its inverse $(k,\omega)$-FT, we obtain%

\begin{equation}
\frac{\partial^{2}\psi_{%
{\scriptscriptstyle{\rm>+}}%
}(x,t)}{\partial x^{2}}-\frac{1}{c^{2}}\frac{\partial^{2}\psi_{%
{\scriptscriptstyle{\rm>+}}%
}(x,t)}{\partial t^{2}}=\frac{m^{2}c^{2}}{\hbar^{2}}\psi_{%
{\scriptscriptstyle{\rm>+}}%
}(x,t), \label{KG0}%
\end{equation}
or, using Einstein's sum convention and the relativistic notation with
$ct\equiv x^{0}$ and $\mu\equiv mc/\hbar,$%

\begin{equation}
(\partial^{\nu}\partial_{\nu}-\mu^{2})\psi_{%
{\scriptscriptstyle{\rm>+}}%
}(x^{\nu})=0. \label{KG0e}%
\end{equation}

Expression (\ref{KG0}) or (\ref{KG0e}) is just the Klein-Gordon (KG) equation.
As it is well known, the KG equation originates from the efforts to fit the
theory of Quantum Mechanics in the relativistic formalism. While we have
obviously used the latter, note that we have derived Eq. (\ref{KG0})
\emph{without} resorting to the usual quantum-mechanical prescriptions:%

\begin{equation}
E\rightarrow i\hbar\partial/\partial t,\qquad p\rightarrow-i\hbar
\partial/\partial x.\label{p2}%
\end{equation}
It must be remarked that Eqs. (\ref{Eh}) and (\ref{ph}), which have also been
used, are simply \textquotedblleft compatible\textquotedblright\ with the
postulates (\ref{p2}) ---but certainly weaker.

We now arrive at the key step. We write $\psi_{%
{\scriptscriptstyle{\rm>+}}%
}(x,t)$ as a product of a \textquotedblleft slow\textquotedblright\ time
complex envelope $\overset{_{\frown}}{\psi}_{%
{\scriptscriptstyle{\rm>}}%
}(x,t)$ and an time exponential oscillatory factor, as in Eq. (\ref{r5}). In
doing so, we chose the frequency of the latter to be precisely $\omega
_{c}=E_{c}/\hbar=mc^{2}/\hbar,$ \emph{i.e.} we associate the Hilbert carrier
to the rest energy of the particle. We shall call $\omega_{c}$ the
\emph{latent pulsation} of $\psi_{%
{\scriptscriptstyle{\rm>+}}%
}(x,t).$

We have%

\begin{equation}
\psi_{%
{\scriptscriptstyle{\rm>+}}%
}(x,t)=\overset{_{\frown}}{\psi}_{%
{\scriptscriptstyle{\rm>}}%
}(x,t)\,e^{-i\omega_{c}t}. \label{pul}%
\end{equation}

It is worth recalling that, as early as in 1925, de Broglie paid attention to
the specific frequency $\nu_{0}=m_{e}c^{2}/h,$ with $m_{e}$ the electron mass,
for which he coined the name \textquotedblleft proper frequency of the
electron\textquotedblright\ \cite{DB}. However, not any kind of relation seems
to have ever been envisioned between such \textquotedblleft proper
frequency\textquotedblright\ and the carrier frequency of an analytical
signal. Operationally, the decomposition (\ref{pul}) is now introduced in the
literature as an ansatz that yields the Schr\"{o}dinger equation as the
nonrelativistic limit of the KG equation (see for example \cite{Greiner}). In
our approach, we will find that $\omega_{c}$ has a relevant meaning when seen
as the \textquotedblleft carrier\textquotedblright\ frequency of an analytical signal.

Replacing Eq. (\ref{pul}) in Eq. (\ref{KG0}), Schr\"{o}dinger's equation is
obtained for $\overset{_{\frown}}{\psi}_{%
{\scriptscriptstyle{\rm>}}%
}(x,t)$:%

\begin{equation}
\frac{\hbar^{2}}{2m}\frac{\partial^{2}\overset{_{\frown}}{\psi}_{%
{\scriptscriptstyle{\rm>}}%
}(x,t)}{\partial x^{2}}=-i\hbar\frac{\partial\overset{_{\frown}}{\psi}_{%
{\scriptscriptstyle{\rm>}}%
}(x,t)}{\partial t}, \label{Sch0}%
\end{equation}
where a term with $\partial^{2}\overset{_{\frown}}{\psi}_{%
{\scriptscriptstyle{\rm>}}%
}(x,t)/\partial t^{2}$ has been neglected in favor of the first derivative,
according to the paraxial approximation \cite{Greiner}. It is the wavefunction
$\psi_{%
{\scriptscriptstyle{\rm>+}}%
}(x,t),$ which incorporates the rest energy, that obeys a covariant equation
determined by the relativity principle, while the slow complex envelope
$\overset{_{\frown}}{\psi}_{%
{\scriptscriptstyle{\rm>}}%
}(x,t)$, which has a displaced $\omega$-spectrum, does not; Schr\"{o}dinger's
wavefunction $\overset{_{\frown}}{\psi}_{%
{\scriptscriptstyle{\rm>}}%
}(x,t)$ is the \textquotedblleft slow\textquotedblright\ time complex envelope
of an analytical signal $\psi_{%
{\scriptscriptstyle{\rm>+}}%
}(x,t)$ with a $\omega$-spectrum located around a very high frequency
$\omega_{c},$ the latent pulsation of the particle.

For a particle in a potential $V(x),$ one considers the space discretized in a
set of subspaces $x_{i}$ and width $\Delta x,$ wherein the potential takes on
constant values $V=V_{i},$ and infers that in the limit the equation valid for
all space encompasses all the local equations and preserves the continuity of
$\psi_{%
{\scriptscriptstyle{\rm>+}}%
}(x,t)$ and its derivative. [This kind of \emph{ad hoc} argument is anything
but appealing theoretically, but in essence it is not so different from the
approaches used in the original derivations (see for example \cite{Sch}).]
Namely, Eqs. (\ref{4}) are now modified to the form%
\begin{equation}
E=mc^{2}(1-v^{2}/c^{2})^{-1/2}+V;\qquad p=mv(1-v^{2}/c^{2})^{-1/2}%
=c^{-1}[(E-V)^{2}-m^{2}c^{4}]^{1/2}. \label{5}%
\end{equation}
Proceeding as previously, there follows the equality%

\begin{equation}
k^{2}=\frac{\omega^{2}}{c^{2}}-\frac{2V}{\hbar c^{2}}\omega+\frac{V^{2}}%
{\hbar^{2}c^{2}}-\frac{m^{2}c^{2}}{\hbar^{2}}. \label{6}%
\end{equation}
One next multiplies Eq. (\ref{6}) by $\tilde{\Psi}_{%
{\scriptscriptstyle{\rm>+}}%
}(k,\omega)$ and takes the inverse $(k,\omega)$-FT, which is still trivial so
long as $V$ is not position-dependent, as assumed. This yields%

\begin{equation}
\frac{\partial^{2}\psi_{%
{\scriptscriptstyle{\rm>+}}%
}(x,t)}{\partial x^{2}}-\frac{1}{c^{2}}\frac{\partial^{2}\psi_{%
{\scriptscriptstyle{\rm>+}}%
}(x,t)}{\partial t^{2}}-i\frac{2V(x)}{\hbar c^{2}}\frac{\partial\psi_{%
{\scriptscriptstyle{\rm>+}}%
}(x,t)}{\partial t}=\left(  \frac{m^{2}c^{2}}{\hbar^{2}}-\frac{V^{2}(x)}%
{\hbar^{2}c^{2}}\right)  \psi_{%
{\scriptscriptstyle{\rm>+}}%
}(x,t), \label{KGx}%
\end{equation}
which generalizes Eq. (\ref{KG0}). Eq. (\ref{KGx}) is the \emph{time-dependent
KG equation with a potential} $V$ (see \cite{Kragh} and references therein).
Of course, as announced, $V$ had to be generalized \emph{ad hoc} in Eq.
(\ref{KGx}) to be position-dependent.

Proceeding analogously to the free particle case, we substitute Eq.
(\ref{pul}) in Eq. (\ref{KGx}) and obtain, when\footnote{Alternatively, Eq.
(\ref{Sch1}) can be derived starting from an approximation of Eq. (\ref{6}) in
the first place. In the non-relativistic limit $v\ll c,$ Eq. (\ref{6}) can be
put, after some manipulations, in the form
\par%
\begin{equation}
k^{2}\simeq\frac{2m}{\hbar}\omega-\left(  \frac{2mV}{\hbar^{2}}-\frac
{2m^{2}c^{2}}{\hbar^{2}}\right)  . \label{7}%
\end{equation}
Multiplying Eq. (\ref{7}) by $\tilde{\Psi}_{%
{\scriptscriptstyle{\rm>+}}%
}(k,\omega)$ and taking its inverse $(k,\omega)$-FT (and setting $V=V(x)$ at
the end), a wave equation is obtained for $\psi_{%
{\scriptscriptstyle{\rm>+}}%
}(x,t)$:
\par%
\begin{equation}
-\frac{\hbar^{2}}{2m}\frac{\partial^{2}\psi_{%
{\scriptscriptstyle{\rm>+}}%
}(x,t)}{\partial x^{2}}-i\hbar\frac{\partial\psi_{%
{\scriptscriptstyle{\rm>+}}%
}(x,t)}{\partial t}+\left(  V+mc^{2}\right)  \psi_{%
{\scriptscriptstyle{\rm>+}}%
}(x,t)=0. \label{KGa}%
\end{equation}
Eq. (\ref{KGa}), which already lacks the term with the second time derivative,
is an approximate version of Eq. (\ref{KGx}) because it has been derived from
the approximate equation (\ref{7}). Using Eq. (\ref{pul}) in Eq. (\ref{KGa}),
Schr\"{o}dinger's equation (\ref{Sch1}) follows exactly.} $V/(mc^{2}%
)\allowbreak\ll\allowbreak1,$ the Schr\"{o}dinger equation \emph{for the time
complex envelope} $\overset{_{\frown}}{\psi}_{%
{\scriptscriptstyle{\rm>}}%
}(x,t)$:%

\begin{equation}
\frac{\hbar^{2}}{2m}\frac{\partial\overset{_{\frown}}{\psi}_{%
{\scriptscriptstyle{\rm>}}%
}(x,t)}{\partial x^{2}}+V(x)\overset{_{\frown}}{\psi}_{%
{\scriptscriptstyle{\rm>}}%
}(x,t)=-i\hbar\frac{\partial\overset{_{\frown}}{\psi}_{%
{\scriptscriptstyle{\rm>}}%
}(x,t)}{\partial t}. \label{Sch1}%
\end{equation}

\section{Average energy\label{s5}}

We are now going to introduce the extremely important concept of \emph{average
temporal frequency} of a \emph{real} wave $\psi(x,t)$. We must first notice
that, $\psi(x,t)$\textbf{\ }being real, $|\Psi(x,\omega)|=|\Psi(x,-\omega)|$,
so it follows that $\int_{-\infty}^{\infty}\omega|\Psi(x,\omega)|^{2}d\omega$
is identically zero. It is then obvious that, if we are to define an average
frequency for any real function, it only makes sense to do it over the
\emph{positive} frequency range. As mentioned before, $\psi_{%
{\scriptscriptstyle{\rm>+}}%
}(x,t)$ is, save for a factor of 2, a signal containing only the positive
$\omega$-frequencies of $\psi(x,t).$ So, we take%

\begin{equation}
\langle\omega\rangle\equiv\frac{\int_{L}dx\int_{-\infty}^{\infty}\omega|\Psi_{%
{\scriptscriptstyle{\rm>+}}%
}(x,\omega)|^{2}d\omega}{\int_{L}dx\int_{-\infty}^{\infty}|\Psi_{%
{\scriptscriptstyle{\rm>+}}%
}(x,\omega)|^{2}d\omega}=N_{E}\int_{-\infty}^{\infty}\omega|\Psi_{%
{\scriptscriptstyle{\rm>+}}%
}(x,\omega)|^{2}d\omega, \label{22}%
\end{equation}
where $L$ is the normalization length (a volume, in three dimensions) and we
have called $N_{E}\equiv\allowbreak\lbrack\int_{L}dx\int_{-\infty}^{\infty
}|\Psi_{%
{\scriptscriptstyle{\rm>+}}%
}(x,\omega)|^{2}d\omega]^{-1}\int_{L}dx\,(\cdot)$ for the sake of notational
brevity. Expression (\ref{22}) would also apply with $\Psi_{+}(x,\omega)$\ in
place of $\Psi_{%
{\scriptscriptstyle{\rm>+}}%
}(x,\omega),$\ but we need to consider the analytical spatial signal as well,
as will be seen in Section \ref{s6}\textbf{.}

We next use Parseval's theorem \cite{Oppenheim}, which states that, for two
arbitrary complex functions $f_{1}(t)$ and $f_{2}(t),$%

\begin{equation}
\int_{-\infty}^{\infty}f_{1}(t)f_{2}^{\ast}(t)dt=\frac{1}{2\pi}\int_{-\infty
}^{\infty}F_{1}(\omega)F_{2}^{\ast}(\omega)d\omega. \label{Par}%
\end{equation}
Taking $F_{1}(\omega)=\omega\Psi_{%
{\scriptscriptstyle{\rm>+}}%
}(x,\omega)$ and $F_{2}(\omega)=\Psi_{%
{\scriptscriptstyle{\rm>+}}%
}(x,\omega)$ in (\ref{22}), and recalling that $f_{1}(t)=\allowbreak$%
FT$^{-1}\{\omega\Psi_{%
{\scriptscriptstyle{\rm>+}}%
}(x,\omega)\}=i\partial\psi_{%
{\scriptscriptstyle{\rm>+}}%
}(x,t)/\partial t,$ we obtain%

\begin{equation}
\langle\omega\rangle=2\pi N_{E}\int_{-\infty}^{\infty}\psi\,_{%
{\scriptscriptstyle{\rm>+}}%
}^{\ast}(x,t)\,i\frac{\partial\psi_{%
{\scriptscriptstyle{\rm>+}}%
}(x,t)}{\partial t}dt. \label{23}%
\end{equation}
Let us write Eq. (\ref{23}) in terms of the temporal complex envelope
$\overset{_{\frown}}{\psi}_{%
{\scriptscriptstyle{\rm>}}%
}(x,t),$ assuming a particle of rest energy $E_{c}$ and latent pulsation
$\omega_{c}=E_{c}/\hbar$. Replacing Eq. (\ref{pul}) in Eq. (\ref{23}), the
following relationship is obtained:%

\begin{equation}
\langle\omega\rangle=\frac{\int_{L}dx\int_{-\infty}^{\infty}\overset{_{\frown
}}{\psi}\,_{%
{\scriptscriptstyle{\rm>}}%
}^{\ast}(x,t)\,\,i\dfrac{\partial\overset{_{\frown}}{\psi}_{%
{\scriptscriptstyle{\rm>}}%
}(x,t)}{\partial t}dt}{\int_{L}dx\int_{-\infty}^{\infty}|\overset{_{\frown}%
}{\psi}_{%
{\scriptscriptstyle{\rm>}}%
}(x,t)|^{2}dt}+\omega_{c}\equiv\langle\overset{_{\frown}}{\omega}%
\rangle+\omega_{c}. \label{Em2}%
\end{equation}
In Eq. (\ref{Em2}) the relationship $\int_{-\infty}^{\infty}|\Psi_{%
{\scriptscriptstyle{\rm>+}}%
}(x,\omega)|^{2}d\omega=2\pi\int_{-\infty}^{\infty}|\overset{_{\frown}}{\psi
}_{%
{\scriptscriptstyle{\rm>}}%
}(x,t)|^{2}dt$ has been used, which can be derived from Eq. (\ref{Par}).

Thus, the average frequency of $\psi_{%
{\scriptscriptstyle{\rm>+}}%
}(x,t)$ is revealed to be the sum of $\omega_{c},$ the latent pulsation of the
rest particle, plus the \emph{average frequency} of the spectrum of its
\textquotedblleft baseband-lying\textquotedblright\ complex envelope, denoted
$\overset{_{\frown}}{\omega}$. Multiplying Eq. (\ref{Em2}) by $\hbar,$ we get
an expression for $\langle E\rangle,$ the \emph{total average energy}, as the
sum of the rest energy of the particle and the average energy of the complex envelope:%

\begin{equation}
\langle E\rangle=\langle\overset{_{\frown}}{E}\rangle+E_{c}, \label{EEc}%
\end{equation}
with%

\begin{equation}
\langle\overset{_{\frown}}{E}\rangle\equiv\hbar\langle\overset{_{\frown}%
}{\omega}\rangle=\frac{\int_{L}dx\int_{-\infty}^{\infty}\overset{_{\frown}%
}{\psi}\,_{%
{\scriptscriptstyle{\rm>}}%
}^{\ast}(x,t)\,i\hbar\dfrac{\partial\overset{_{\frown}}{\psi}_{%
{\scriptscriptstyle{\rm>}}%
}(x,t)}{\partial t}dt}{\int_{L}dx\int_{-\infty}^{\infty}|\overset{_{\frown}%
}{\psi}_{%
{\scriptscriptstyle{\rm>}}%
}(x,t)|^{2}dt}. \label{Em3}%
\end{equation}

As we saw in Section \ref{s3and4}, it is the wavefunction $\overset{_{\frown}%
}{\psi}_{%
{\scriptscriptstyle{\rm>}}%
}(x,t)$ which appears in the Schr\"{o}dinger equation. Consequently, we have
been able to derive the expression (\ref{Em3}) for the average value of the
energy without using the quantum-mechanical postulates (\ref{p2}). In our
approach, based only on the formula (\ref{Eh}), Eq. (\ref{Em3}) is obtained in
a natural way. In fact, calling%

\begin{equation}
\hat{E}\equiv i\hbar\frac{\partial}{\partial t}, \label{defE}%
\end{equation}
we can write, for the \emph{average energy},%

\begin{equation}
\langle\overset{_{\frown}}{E}\rangle=\frac{\int_{L}dx\int_{-\infty}^{\infty
}\overset{_{\frown}}{\psi}\,_{%
{\scriptscriptstyle{\rm>}}%
}^{\ast}(x,t)\hat{E}\overset{_{\frown}}{\psi}_{%
{\scriptscriptstyle{\rm>}}%
}(x,t)dt}{\int_{L}dx\int_{-\infty}^{\infty}|\overset{_{\frown}}{\psi}_{%
{\scriptscriptstyle{\rm>}}%
}(x,t)|^{2}dt}. \label{Em4}%
\end{equation}

The notation (\ref{defE}) is meant to indicate that $i\hbar\partial/\partial
t$ is the \textquotedblleft energy quantum operator,\textquotedblright\ as the
result (\ref{Em4}) strongly suggests. However, Eq. (\ref{Em4}) is not a
sufficient condition. We will return to this point in Section \ref{s7}.

Note also that if the integration over all space, contained in $N_{E},$ had
not been performed, $\langle E\rangle$ would have been $x$-dependent, which
diverts from the usual consideration of the energy as a global, non-localized,
characteristic of the system. Finally, it is important to realize that all the
formalism developed in this section is virtually \emph{independent of the form
of the wave equation obeyed} by $\overset{_{\frown}}{\psi}_{%
{\scriptscriptstyle{\rm>}}%
}.$

\section{Schr\"{o}dinger's time-independent equation\label{sest}}

In looking for pure time harmonic solutions, one finds that the so-called
Schr\"{o}dinger's time-independent equation is essentially connected to the
$\omega$-FT of Eq. (\ref{Sch1}). The latter reads, denoting $\overset
{_{\frown}}{\omega}=\overset{_{\frown}}{E}/\hbar$ the transformed variable,%

\begin{equation}
-\frac{\hbar^{2}}{2m}\frac{\partial^{2}\overset{_{\frown}}{\Psi}_{%
{\scriptscriptstyle{\rm>}}%
}(x,\overset{_{\frown}}{\omega})}{\partial x^{2}}+V(x)\overset{_{\frown}}%
{\Psi}_{%
{\scriptscriptstyle{\rm>}}%
}(x,\overset{_{\frown}}{\omega})=\overset{_{\frown}}{E}\overset{_{\frown}%
}{\Psi}_{%
{\scriptscriptstyle{\rm>}}%
}(x,\overset{_{\frown}}{\omega}). \label{st}%
\end{equation}
Such Fourier relation is straightforward, but is usually overlooked due to the
fact that a \emph{discrete} sum of complex time exponential corresponds to a
sum of Dirac deltas in the frequency domain. Properly, one has to distinguish
two cases:

(1) If $E>V(x),$ then the solutions of (\ref{st}) can be simply written as%

\begin{equation}
\overset{_{\frown}}{\Psi}_{%
{\scriptscriptstyle{\rm>}}%
\,\overset{_{\frown}}{E}}(x,\overset{_{\frown}}{\omega})\equiv\overset
{_{\frown}}{\varphi}_{%
{\scriptscriptstyle{\rm>}}%
\,\overset{_{\frown}}{E}}(x),
\end{equation}
with $\overset{_{\frown}}{E}=\hbar\overset{_{\frown}}{\omega},$ $\overset
{_{\frown}}{\varphi}_{%
{\scriptscriptstyle{\rm>}}%
\,\overset{_{\frown}}{E}}(x)$ being the complex function conveying the spatial
information of the wavefunction associated to the energy $\overset{_{\frown}%
}{E},$ which can have any value in a continuous range. With this more familiar
notation, the time-independent Schr\"{o}dinger equation reads%

\begin{equation}
-\frac{\hbar^{2}}{2m}\frac{\partial^{2}\overset{_{\frown}}{\varphi}_{%
{\scriptscriptstyle{\rm>}}%
\,\overset{_{\frown}}{E}}(x)}{\partial x^{2}}+V(x)\overset{_{\frown}}{\varphi
}_{%
{\scriptscriptstyle{\rm>}}%
\,\overset{_{\frown}}{E}}(x)=\overset{_{\frown}}{E}\overset{_{\frown}}%
{\varphi}_{%
{\scriptscriptstyle{\rm>}}%
\,\overset{_{\frown}}{E}}(x).
\end{equation}

These continuous-frequency solutions are typically found in cases such as the
propagation in infinite periodic media (Bloch waves), unbounded quantum-well
solutions, etc. They form a representation base for wavefunctions with a
continuous energy spectrum, so that%

\begin{equation}
\int_{-\infty}^{\infty}\overset{_{\frown}}{\varphi}\,_{%
{\scriptscriptstyle{\rm>}}%
\,\overset{_{\frown}}{E}}^{\ast}(x)\overset{_{\frown}}{\varphi}_{%
{\scriptscriptstyle{\rm>}}%
\,\overset{_{\frown}}{E}^{\prime}}(x)dx=\delta(\overset{_{\frown}}{E}%
-\overset{_{\frown}}{E}^{\prime}).
\end{equation}

(2) If $E<V(x),$ one has discrete eigenfunctions of (\ref{st}), that need be
written as%

\begin{equation}
\overset{_{\frown}}{\Psi}_{%
{\scriptscriptstyle{\rm>}}%
\,n}(x,\overset{_{\frown}}{\omega}_{n})\equiv\overset{_{\frown}}{\varphi}_{%
{\scriptscriptstyle{\rm>}}%
\,n}(x)\,\delta(\overset{_{\frown}}{\omega}-\overset{_{\frown}}{\omega}_{n}),
\end{equation}
the corresponding eigenvalues being $\overset{_{\frown}}{E}_{n}=\hbar
\overset{_{\frown}}{\omega}_{n}.$ Then, there follows%

\begin{equation}
-\frac{\hbar^{2}}{2m}\frac{\partial^{2}\overset{_{\frown}}{\varphi}_{%
{\scriptscriptstyle{\rm>}}%
\,n}(x)}{\partial x^{2}}+V(x)\overset{_{\frown}}{\varphi}_{%
{\scriptscriptstyle{\rm>}}%
\,n}(x)=\overset{_{\frown}}{E}\overset{_{\frown}}{\varphi}_{%
{\scriptscriptstyle{\rm>}}%
\,n}(x). \label{42}%
\end{equation}

It is this case that corresponds to the true \emph{stationary} waves, made up
of pairs of identical waves propagating in both directions, $>$ and $<.$ In
fact, since (\ref{42}) also holds for $\overset{_{\frown}}{\varphi}_{%
{\scriptscriptstyle{\rm<}}%
\,n}(x)=\overset{_{\frown}}{\varphi}_{%
{\scriptscriptstyle{\rm>}}%
\,n}^{\,\ast}(x),$ adding both equations yields%

\begin{equation}
-\frac{\hbar^{2}}{2m}\frac{\partial^{2}\overset{_{\frown}}{\varphi}_{n}%
(x)}{\partial x^{2}}+V(x)\overset{_{\frown}}{\varphi}_{\,n}(x)=\overset
{_{\frown}}{E}_{n}\overset{_{\frown}}{\varphi}_{n}(x),\label{tst}%
\end{equation}
with is the true time-independent \emph{stationary} Schr\"{o}dinger's
equation. Eq. (\ref{tst}) is the wave equation that applies to one-dimensional
cases such as an infinite potential well, the bounded wavefunctions of a
finite potential well, the harmonic oscillator, etc.\ Indeed, it is well
known, and can be proved algebraically with no difficulty, that the
eigenfunctions corresponding to discrete non-degenerate eigenvalues are
necessarily \emph{real-valued,} the mentioned cases being typical examples.

The corresponding Schr\"{o}dinger $n$-eigenfunctions are%

\begin{equation}
\overset{_{\frown}}{\psi}_{%
{\scriptscriptstyle{\rm>}}%
\,n}(x,t)=\overset{_{\frown}}{\varphi}_{%
{\scriptscriptstyle{\rm>}}%
\,n}(x)\,e^{-i\overset{_{\frown}}{E}_{n}t/\hbar}. \label{base}%
\end{equation}

Now recalling Eq. (\ref{pul}), the following result is obtained:%

\begin{equation}
\psi_{n\,%
{\scriptscriptstyle{\rm>+}}%
}(x,t)=\overset{_{\frown}}{\varphi}_{%
{\scriptscriptstyle{\rm>}}%
\,n}(x)\,e^{-i\overset{_{\frown}}{E}_{n}t/\hbar}e^{-iE_{c}t/\hbar}%
\equiv\overset{_{\frown}}{\varphi}_{%
{\scriptscriptstyle{\rm>}}%
\,n}(x)\,e^{-iE_{n}t/\hbar}.
\end{equation}

We thus find that the oscillation frequencies have the form $\omega_{n}%
=\omega_{c}+\overset{_{\frown}}{\omega}_{n}=E_{c/}\hbar+\overset{_{\frown}}%
{E}_{n}/\hbar$. This is, each $\omega_{n}$ contains the rest energy $E_{c}$ in
addition to the familiar energy eigenvalues of the Schr\"{o}dinger equation,
$\overset{_{\frown}}{E}_{n}.$ [This is, naturally, in agreement with the
result (\ref{EEc}).]

\section{Average momentum\label{s6}}

As in Section \ref{s5}, we expect that, in view of Eq. (\ref{ph}), the average
momentum in the quantum state $\psi(x,t)$ can be obtained from the average
wavevector of the wavefunction, $\langle k\rangle.$ As is the case of the
temporal frequency $\omega$, the average $k$ of any real-valued signal is
strictly zero, so it only makes sense to compute $\langle k\rangle$ with the
spatial \emph{analytical} signal. In fact, we have already adopted this choice
with $\psi_{%
{\scriptscriptstyle{\rm>+}}%
}(x,t)$. We can write%

\begin{equation}
\langle k\rangle\equiv\frac{\int_{-\infty}^{\infty}k|\tilde{\psi}_{%
{\scriptscriptstyle{\rm>+}}%
}(k,t)|^{2}dk}{\int_{-\infty}^{\infty}|\tilde{\psi}_{%
{\scriptscriptstyle{\rm>+}}%
}(k,t)|^{2}dk}=\frac{\int_{-\infty}^{\infty}k|\tilde{\overset{_{\frown}}{\psi
}}\,_{%
{\scriptscriptstyle{\rm>}}%
}(k,t)|^{2}dk}{\int_{-\infty}^{\infty}|\tilde{\overset{_{\frown}}{\psi}}_{%
{\scriptscriptstyle{\rm>}}%
}(k,t)|^{2}dk}\equiv N_{k}\int_{-\infty}^{\infty}k\,\tilde{|\overset{_{\frown
}}{\psi}}_{%
{\scriptscriptstyle{\rm>}}%
}(k,t)|^{2}dk, \label{24}%
\end{equation}
with $N_{k}\equiv\lbrack\int_{-\infty}^{\infty}|\tilde{\overset{_{\frown}%
}{\psi}}_{%
{\scriptscriptstyle{\rm>}}%
}(k,t)|^{2}dk]^{-1}.$ In Eq. (\ref{24}) we have chosen to express $\langle
k\rangle$ in terms of the complex time envelope $\tilde{\overset{_{\frown}%
}{\psi}}_{%
{\scriptscriptstyle{\rm>}}%
}(k,t),$ rather than the analytical time signal $\tilde{\psi}_{%
{\scriptscriptstyle{\rm>+}}%
}(k,t),$ because we found in Section \ref{s3and4} that it is the former that
appears in the standard Schr\"{o}dinger equation (\ref{Sch1}). Unlike the
$\omega$-spectrum, there is no Hilbert frequency in the space domain since the
momentum of the rest particle is zero. Consequently, the $k$ spectrum of
$\psi(x,t)$ is \textquotedblleft low band\textquotedblright\ and there is no
need to consider any factorization analogous to Eq. (\ref{pul}).

We use again Parseval's theorem (\ref{22}) with $t\rightarrow x$ and
$\omega\rightarrow k,$ applied to $F_{1}(k)=k\tilde{\overset{_{\frown}}{\psi}%
}_{%
{\scriptscriptstyle{\rm>}}%
}(k,t)$ and $F_{2}(k)=\tilde{\overset{_{\frown}}{\psi}}_{%
{\scriptscriptstyle{\rm>}}%
}(k,t)$. Using $f_{1}(x)=k$-FT$^{-1}\{k\tilde{\overset{_{\frown}}{\psi}}_{%
{\scriptscriptstyle{\rm>}}%
}(k,t)\}=-i\partial\overset{_{\frown}}{\psi}_{%
{\scriptscriptstyle{\rm>}}%
}(x,t)/\partial x,$ we get%

\begin{equation}
\langle k\rangle=\frac{-\int_{-\infty}^{\infty}\overset{_{\frown}}{\psi}\,_{%
{\scriptscriptstyle{\rm>}}%
}^{\ast}(x,t)\,i\dfrac{\partial\overset{_{\frown}}{\psi}_{%
{\scriptscriptstyle{\rm>}}%
}(x,t)}{\partial x}dx}{\int_{-\infty}^{\infty}|\overset{_{\frown}}{\psi}_{%
{\scriptscriptstyle{\rm>}}%
}(x,t)|^{2}dk}. \label{pm1}%
\end{equation}

Multiplying Eq. (\ref{pm1}) by $\hbar$, using Eq. (\ref{ph}), and calling%

\begin{equation}
\hat{p}\equiv-i\hbar\frac{\partial}{\partial x}, \label{pop}%
\end{equation}
we can write, for the \emph{average momentum},%

\begin{equation}
\langle p\rangle=\frac{\int_{-\infty}^{\infty}\overset{_{\frown}}{\psi}\,_{%
{\scriptscriptstyle{\rm>}}%
}^{\ast}(x,t)\,\,\hat{p}\,\overset{_{\frown}}{\psi}_{%
{\scriptscriptstyle{\rm>}}%
}(x,t)dx}{\int_{-\infty}^{\infty}|\overset{_{\frown}}{\psi}_{%
{\scriptscriptstyle{\rm>}}%
}(x,t)|^{2}dk}. \label{pp}%
\end{equation}
Eq. (\ref{pp}) has the same form as the standard expression for the average
value of the quantum-mechanical momentum when the system is in the state
$\overset{_{\frown}}{\psi}_{%
{\scriptscriptstyle{\rm>}}%
}(x,t).$ Once again, this result has been derived without resorting to the
postulates (\ref{p2}). That $-i\hbar\partial/\partial x$ is truly the
\textquotedblleft quantum momentum operator\textquotedblright\ will be seen in
the next section.

\section{Measurement probability\label{s7}}

In Sections \ref{s5} and \ref{s6} we have obtained the expressions of the
average energy and momentum, Eqs. (\ref{Em3}) and (\ref{pp}), respectively.
Although these results are promising, the much more general probability
postulate remains to be justified. With this purpose, we consider the
Schr\"{o}dinger wavefunction of an arbitrary quantum state, expanded in the
base of stationary Schr\"{o}dinger $n$-eigenfunctions, assumed discrete, Eq.
(\ref{base}):%

\begin{equation}
\overset{_{\frown}}{\psi}_{%
{\scriptscriptstyle{\rm>}}%
}(x,t)=\sum_{n}a_{n}\overset{_{\frown}}{\varphi}_{%
{\scriptscriptstyle{\rm>}}%
\,n}(x)\,e^{-i\overset{_{\frown}}{E}_{n}t/\hbar}, \label{uu}%
\end{equation}

Making use of the formalism developed in Section \ref{s5}, the $r$-th moment
of the complex envelope frequency is found to be%

\begin{equation}
\langle\overset{_{\frown}}{\omega}^{r}\rangle=\frac{\int_{L}dx\int_{-\infty
}^{\infty}\overset{_{\frown}}{\Psi}_{%
{\scriptscriptstyle{\rm>}}%
}^{\ast}(x,\omega)\omega^{r}\overset{_{\frown}}{\Psi}_{%
{\scriptscriptstyle{\rm>}}%
}(x,\omega)d\omega}{\int_{L}dx\int_{-\infty}^{\infty}|\overset{_{\frown}}%
{\Psi}_{%
{\scriptscriptstyle{\rm>}}%
}(x,\omega)|^{2}d\omega}. \label{wr}%
\end{equation}
Using Eq. (\ref{Par}), recalling the relation $\omega^{r}\Psi(x,\omega
)\leftrightarrow i^{r}\partial^{r}\psi(x,t)/\partial t^{r},$ and multiplying
Eq. (\ref{wr}) by $\hbar^{r},$ we obtain%

\begin{equation}
\langle\overset{_{\frown}}{E}^{r}\rangle=\hbar^{r}\langle\overset{_{\frown}%
}{\omega}^{r}\rangle=\frac{\int_{L}dx\int_{-\infty}^{\infty}\overset{_{\frown
}}{\psi}_{%
{\scriptscriptstyle{\rm>}}%
}^{\ast}(x,t)\,(i\hbar)^{r}\dfrac{\partial^{r}\overset{_{\frown}}{\psi}_{%
{\scriptscriptstyle{\rm>}}%
}(x,t)}{\partial t^{r}}dt}{\int_{L}dx\int_{-\infty}^{\infty}|\overset
{_{\frown}}{\psi}_{%
{\scriptscriptstyle{\rm>}}%
}(x,t)|^{2}dt}. \label{Er1}%
\end{equation}
Using the expansion in the base (\ref{uu}), the orthonormality condition
$\int_{L}\overset{_{\frown}}{\varphi}_{%
{\scriptscriptstyle{\rm>}}%
\,n}^{\ast}(x)\overset{_{\frown}}{\varphi}_{%
{\scriptscriptstyle{\rm>}}%
\,m}(x)dx=\delta_{nm}$ and the normalization $\sum_{n}|a_{n}|^{2}=1,$ we obtain%

\begin{equation}
\langle\overset{_{\frown}}{E}^{r}\rangle=\hbar^{r}\langle\overset{_{\frown}%
}{\omega}^{r}\rangle=\sum_{n}|a_{n}|^{2}\overset{_{\frown}}{E}_{n}^{r}.
\label{momr}%
\end{equation}

It is obvious that, \emph{if} $|a_{n}|^{2}$ were the probability of measuring
the energy $\overset{_{\frown}}{E}_{n},$ then the result (\ref{momr}) would
immediately follow. However, it is \emph{the sufficient condition} that we
must prove; i.e., that, if the moment formula (\ref{momr}) holds, then
$|a_{n}|^{2}$ is the \emph{probability} of measuring the energy $\overset
{_{\frown}}{E}_{n}$ (above $E_{c},$ actually) when the particle is in the
state $\overset{_{\frown}}{\psi}_{%
{\scriptscriptstyle{\rm>}}%
}(x,t).$ This is indeed so and we give the proof next.

In order to prove that $|a_{n}|^{2}$ is the probability of measuring
$\overset{_{\frown}}{E}_{n}$---or, in the continuous case, that $|a(\overset
{_{\frown}}{E})|^{2}$ is the probability density for $\overset{_{\frown}}{E}%
$---, we will make use of the so-called \emph{characteristic function} of a
random variable, which, for a continuous variable with probability density
$P(\overset{_{\frown}}{E}),$ is defined as \cite{Papoulis}%

\begin{equation}
G(s)\equiv\int_{-\infty}^{\infty}P(\overset{_{\frown}}{E})e^{is\overset
{_{\frown}}{E}}dE. \label{a2}%
\end{equation}
Now, $\exp(is\overset{_{\frown}}{E})=\sum_{r=0}^{\infty}(is)^{r}%
\overset{_{\frown}}{E}^{r}/r!,$ which, replaced in (\ref{a2}) yields%

\begin{equation}
G(s)=\sum_{r=0}^{\infty}\frac{(is)^{r}}{r!}\langle\overset{_{\frown}}{E}%
^{r}\rangle. \label{a4}%
\end{equation}
For a discrete probability distribution, the characteristic function is
defined as $G(s)\equiv\sum_{n}P_{n}e^{isE_{n}},$ and the same result
(\ref{a4}) is obtained.

In our case, we have%

\begin{equation}
G(s)=\sum_{r=0}^{\infty}\frac{(is)^{r}}{r!}\langle\overset{_{\frown}}{E}%
^{r}\rangle=\left\{
\begin{array}
[c]{c}%
{\displaystyle\sum\limits_{r=0}^{\infty}}
\dfrac{(is)^{r}}{r!}%
{\displaystyle\sum\limits_{n}}
|a_{n}|^{2}\overset{_{\frown}}{E}_{n}^{r}\qquad\text{(discrete)}\\%
{\displaystyle\sum\limits_{r=0}^{\infty}}
\dfrac{(is)^{r}}{r!}%
{\displaystyle\int\limits_{\overset{_{\frown}}{E}}}
|a(\overset{_{\frown}}{E})|^{2}\overset{_{\frown}}{E}^{r}d\overset{_{\frown}%
}{E}^{r}\qquad\text{(continuous).}%
\end{array}
\right.  \label{8}%
\end{equation}
For a continuous distribution, the inverse transform of (\ref{a2}) then reads%

\begin{align}
P(\overset{_{\frown}}{E}) &  \equiv\frac{1}{2\pi}\int_{-\infty}^{\infty
}G(s)e^{-is\overset{_{\frown}}{E}}ds=\frac{1}{2\pi}\int_{-\infty}^{\infty}%
\sum_{r=0}^{\infty}\frac{(is)^{r}}{r!}\left[  \int_{\overset{_{\frown}}{E}%
}|a(\overset{_{\frown}}{E})|^{2}\overset{_{\frown}}{E}^{r}d\overset{_{\frown}%
}{E}\right]  e^{-is\overset{_{\frown}}{E}}ds\nonumber\\
&  =\frac{1}{2\pi}\int_{\overset{_{\frown}}{E}^{\prime}}d\overset{_{\frown}%
}{E}^{\prime}|a(\overset{_{\frown}}{E}^{\prime})|^{2}\int_{-\infty}^{\infty
}e^{-is\overset{_{\frown}}{E}}\underset{\exp(is\overset{_{\frown}}{E}^{\prime
})}{\underbrace{\sum_{r=0}^{\infty}\frac{1}{r!}\overset{_{\frown}}{E}^{\prime
r}(is)^{r}}}ds\nonumber\\
&  =\frac{1}{2\pi}\int_{\overset{_{\frown}}{E}^{\prime}}d\overset{_{\frown}%
}{E}^{\prime}|a(\overset{_{\frown}}{E}^{\prime})|^{2}2\pi\delta(\overset
{_{\frown}}{E}^{\prime}-\overset{_{\frown}}{E})=|a(\overset{_{\frown}}%
{E})|^{2},\text{\quad Q.E.D.}%
\end{align}
The discrete distribution can be treated in the same framework by putting
$|a(\overset{_{\frown}}{E})|^{2}=\sum_{n}|a_{n}|^{2}\delta(\overset{_{\frown}%
}{E}-\overset{_{\frown}}{E}_{n}).$ The corresponding result is obtained straightforwardly.

It is interesting to note that, in the derivation of expression (\ref{momr}),
the quotient $\int_{-\infty}^{\infty}dt\,/\allowbreak\int_{-\infty}^{\infty
}dt$ arises, leading to a mathematical difficulty. However, since the result
(\ref{momr}) is known to be \emph{correct}, it appears we should decide that
$\int_{-\infty}^{\infty}dt\,/\int_{-\infty}^{\infty}dt=1$. This problem indeed
resembles that of the normalization of the wavefunction of a free particle in
an infinite volume. In the present case, we may argue that the integrals can
be thought of as to be extended over the interval $[-T,T],$ with $T$
\emph{very large} (but not infinite) as $\overset{_{\frown}}{\psi}_{%
{\scriptscriptstyle{\rm>}}%
}(x,t)$ is $0$ for all practical purposes at remote times.

Again, we have derived this quantum-mechanical postulate by using only Eq.
(\ref{ph}). It is also clear that $i\hbar\partial/\partial t$ should be
considered as the energy operator, as we conjectured in Section \ref{s5}.
Naturally, the quantum-mechanical formula for the average energy is the
particular case of Eq. (\ref{momr}) with $r=1.$

It is trivial to carry out a similar derivation for a continuous energy spectrum,%

\begin{equation}
\overset{_{\frown}}{\psi}_{%
{\scriptscriptstyle{\rm>}}%
}(x,t)=\int_{\overset{_{\frown}}{E}}a(\overset{_{\frown}}{E})\overset
{_{\frown}}{\varphi}_{%
{\scriptscriptstyle{\rm>}}%
\,\overset{_{\frown}}{E}}(x)\,e^{-i\overset{_{\frown}}{E}t/\hbar}%
d\overset{_{\frown}}{E},
\end{equation}
with $\int_{L}\overset{_{\frown}}{\varphi}_{%
{\scriptscriptstyle{\rm>}}%
\,\overset{_{\frown}}{E}}^{\ast}(x)\overset{_{\frown}}{\varphi}_{%
{\scriptscriptstyle{\rm>}}%
\,\overset{_{\frown}}{E}\,^{\prime}}^{{}}(x)dx=\delta(\overset{_{\frown}}%
{E}\mathcal{-}\overset{_{\frown}}{E}\,^{\prime}).$

The derivation for the momentum follows the same guidelines. Let $p_{j}=\hbar
k_{j}$ be a set of momentum eingenvalues. We write%

\begin{equation}
\overset{_{\frown}}{\psi}_{%
{\scriptscriptstyle{\rm>}}%
}(x,t)=\sum_{j}b_{j}(t)e^{ik_{j}x}.
\end{equation}
Generalizing Eq. (\ref{pm1}), we have%

\begin{align}
\langle p^{r}\rangle &  =\hbar^{r}\langle k^{r}\rangle=\hbar^{r}\frac
{\int_{-\infty}^{\infty}k^{r}|\tilde{\overset{_{\frown}}{\psi}}\,_{%
{\scriptscriptstyle{\rm>}}%
}(k,t)|^{2}dk}{\int_{-\infty}^{\infty}|\tilde{\overset{_{\frown}}{\psi}}\,_{%
{\scriptscriptstyle{\rm>}}%
}(k,t)|^{2}dk}=\frac{\int_{-\infty}^{\infty}\overset{_{\frown}}{\psi}\,_{%
{\scriptscriptstyle{\rm>}}%
}^{\ast}(x,t)\,(-i\hbar)^{r}\dfrac{\partial^{r}\overset{_{\frown}}{\psi}_{%
{\scriptscriptstyle{\rm>}}%
}(x,t)}{\partial x^{r}}dx}{\int_{-\infty}^{\infty}|\overset{_{\frown}}{\psi}_{%
{\scriptscriptstyle{\rm>}}%
}(x,t)|^{2}dx}\nonumber\\
&  =\sum_{j}|\,b_{j}(t)|^{2}(\hbar k_{j})^{r},
\end{align}
so $|\,b_{j}(t)|^{2}$ is the probability of measuring the momentum
$p_{j}=\hbar k_{j}$ and $i\hbar\partial/\partial x$ is found to be the
momentum operator, as presumed.

As a final point, we see that the observation made at the end of Section
\ref{s5} is confirmed: The formalism that allows to derive the form of the
quantum operators and the probability formulas for the momentum and energy,
relies purely on the Fourier theory and statistics. Although the specific
\textquotedblleft eigenstates of the Schr\"{o}dinger
equation,\textquotedblright\ for example, have been brought up in the
discussion, a quick review of the procedure reveals that the derivation is not
really subordinated, at a deep level, to the \emph{specific form} of the wave equation.

\section{Concluding remarks\label{s8}}

As it has been seen in the preceding sections, the standard symbol
\textquotedblleft$\psi(x,t)$\textquotedblright\ for the quantum wavefunction
appears very scarcely in this article. On the contrary, the unfamiliar and
cumbersome notations $\overset{_{\frown}}{\psi}_{%
{\scriptscriptstyle{\rm>}}%
}(x,t)$ and $\psi_{%
{\scriptscriptstyle{\rm>+}}%
}(x,t)$ have been used abundantly. Indeed, this deliberate typographical waste
has been the price to pay to keep clear at all times what the so-called
Schr\"{o}dinger's wavefunction is, avoiding any confussion with other related
but different functions.

The intriguing resemblance of the analytical signal theory with the formalism
of Schr\"{o}dinger's wavefunction was the clue that motivated this work. The
form of Schr\"{o}dinger's complex wavefunction, $\overset{_{\frown}}{\psi}_{%
{\scriptscriptstyle{\rm>}}%
}(x,t)$ in our notation, suggested that it might in fact be the temporal
analytical part of some real-valued space-time function containing a rapid
oscillation. We have denoted $\psi_{%
{\scriptscriptstyle{\rm>+}}%
}(x,t)$ its corresponding analytical signal, whose Hilbert carrier is the
ultrafast oscillation corresponding to the rest energy of the particle.

As far as Quantum mechanics is concerned, we have started from the two very
basic relationships (\ref{Eh}) and (\ref{ph}) that associate a particle to a
wave. Other than that, we have only applied wave theory is strict terms,
without resorting to any other quantum postulate.

Note that nothing indicated \emph{a priori} that these functions should be
spatially analytical as well, since the equations would equally apply to the
corresponding functions without the $>$ subscript. However, in order to prove
the momentum postulate (Section \ref{s6}), we have anticipated from the very
start that the complex wavefunctions need also be analytical in the spatial spectrum.

Using wave mechanics exclusively, we have been able to give a plausible
explanation for the fundamental expressions of the energy and momentum quantum
operators, Eq. (\ref{p2}) (only for the one-particle case). To accomplish
this, we have first dealt with the computation of the average values, and then
derived the general postulate of the measurement probability for the energy
and the momentum. This simultaneously yielded the form of the operators.
However, of course, no light can be shed on the part of the postulate
concerning \emph{the collapse of the wavefunction}.

In this work, no attempt has certainly been made to deal, for example, with
multiparticle systems, continuous systems (field quantization) or
Hamiltonian-Lagrangian approaches. We have focused on a simple quantum system,
and a collection of surprising results have been obtained by simply looking at
the complex wavefunction from a fresh perspective. Some non-relativistic
applications of the formalism and further development of the theory, including
the phase problem, will be the object of future work.


\begin{thebibliography}{99}                                                                                               %


\bibitem {Field}Field J H 2004 Relationship of quantum mechanics to classical
electromagnetism and classical relativity mechanics\ \emph{Eur. J. Phys.
}\textbf{25} 385-97

\bibitem {Javier}Fraile-Pelaez, F J 2003 Analytical signal formalism in the
description of optical pulse photodetection\ \emph{Microwave and Opt. Tech.
Lett.} \textbf{37} 347-52

\bibitem {CohenT}Cohen-Tannoudji C, Dupont-Roc J and Grynberg G 1989
\emph{Photons and Atoms - Introduction to Quantum Electrodynamics} (New York:
Wiley) pp. 79--125

\bibitem {Loudon}Loudon R 2000 \emph{The Quantum Theory of Light}\ 3rd ed.
(Oxford: Oxford University Press) pp. 125--147

\bibitem {Adler}Adler C G 1987 Does mass really depend on velocity,
dad?\ \emph{Am. J. Phys.} \textbf{55} 739-43

\bibitem {DB}de Broglie L 1925 Sur la fr\'{e}quence propre de
l'\'{e}lectron\ \emph{Comptes Rendus} \textbf{180} 498-500

\bibitem {Greiner}Greiner W 2000 \emph{Relativistic Quantum Mechanics} 3rd ed.
(Berlin:\ Springer) pp. 7-8

\bibitem {Sch}E. Schr\"{o}dinger, \textquotedblleft An undulatory theory of
the mechanics of atoms and molecules,\textquotedblright\ \emph{Phys. Rev.}
\textbf{28} 1049-70

\bibitem {Kragh}Kragh H 1984 Equations with many fathers. The Klein-Gordon
equation in 1928\ \emph{Am. J. Phys.} \textbf{52} 1024-1033

\bibitem {Oppenheim}Oppenheim A V, Willsky A S, with Nawab S H 1997
\emph{Signals and Systems}\ (Englewood Cliffs, NJ: Prentice-Hall) pp. 211--12

\bibitem {Papoulis}Papoulis A and Unnikrishna Pillai S 2002 \emph{Probability,
Random Variables and Stochastic Processes}\ 4th edn (New York: McGraw-Hill)
pp. 123--68
\end{thebibliography}
\end{document}